\documentclass[aps,prc,preprint,groupedaddress]{revtex4}
\def\ohalf{{\textstyle{1\over 2}}}

\def\thalf{{\textstyle{3\over 2}}}
\def\tturd{{\textstyle{2\over 3}}}
\def\one{{\rm{1}}}
\newcommand{\beq}{\begin{equation}}
\newcommand{\eeq}{\end{equation}}
\newcommand{\beqa}{\begin{eqnarray}}
\newcommand{\eeqa}{\end{eqnarray}}
\newcommand{\bra}[1]{\langle {#1} |}                        % < |
\newcommand{\ket}[1]{| {#1} \rangle}
\newcommand{\clg}[6]
                {\langle #1,#2,#3,#4 ; #5 #6 \rangle}
\def\A{{\cal A}}
\def\M{{\cal M}}
\def\sP{{\sf P}}
\def\sR{{\sf R}}
\def\P{{\cal P}}

\def\L{{\cal L}}

\usepackage{epsfig}

\begin{document}

\title{ Axial Transition Form Factors and Pion Decay of Baryon Resonances }

\author{B. Juli\'a-D\'{\i}az}
\email[]{Bruno.Julia@helsinki.fi}

\author{D. O. Riska}
\email[]{riska@pcu.helsinki.fi}
\affiliation{Helsinki Institute of Physics
and Department of Physical Sciences, POB 64,
00014 University of Helsinki, Finland}

\author{F. Coester}
\email[]{coester@anl.gov}
\affiliation{Physics Division, Argonne National Laboratory,
Argonne, IL 60439, USA}

\date{\today}
\begin{abstract}

The pion decay constants of the lowest orbitally excited 
states of the nucleon and the $\Delta(1232)$ along with 
the corresponding axial transition form factors are calculated 
with Poincar\'e covariant constituent-quark models with
instant, point and front forms of relativistic kinematics.
The model wave functions are chosen such that the calculated 
electromagnetic and axial form factors of the nucleon represent
the empirical values in all three forms of kinematics, 
when calculated with single-constituent currents. 
The pion decay widths calculated with the three forms 
of kinematics are smaller than the empirical values. 
Front and instant form kinematics provide a similar 
description, with a slight preference for front form, 
while the point form values are significantly smaller
in the case of the lowest positive parity resonances.

\end{abstract}

\pacs{}

\maketitle

\section{Introduction}
\label{intro}

Quark model calculations of the pion decay widths of the baryon
resonances typically fail to agree with values that have been 
extracted from pion-nucleon scattering data, without preference 
for any particular Hamiltonian model~\cite{Cano, Capstick, Theussl}. This is 
already evident in the static quark model, in which the value for the 
$\pi N\Delta(1232)$ coupling leads to a $\sim 40\%$ 
underprediction of the decay width for the $\Delta(1232)$ 
resonance~\cite{Hemmert}. The likely reason for this is the
restricted Hilbert space of the constituent-quark model without additional
quark-antiquark configurations~\cite{Lee}. The problem of the pion 
decay widths may also reflect unrealistic features of 
the model wave functions and/or the associated 
axial quark currents. The purpose of this paper is to address this 
issue by a comparison of the effect of different quark currents 
generated by the dynamics from kinematic single-quark currents.

A relativistic constituent-quark phenomenology of baryon properties 
can be implemented with simple spectral representations of the mass operator.
The 
Poincar\'e generators are functions of the mass and spin operators as well as
kinematic quantities, which depend on the choice of a kinematic sub-group 
(the ``form of kinematics''). Vector and axial vector currents are 
generated by the dynamics from single-quark currents.
Single-quark current matrices are by definition functions of kinematic quark
momenta
and spins, which are related to internal momenta and constituent spins 
by boost transformations that depend on the form of kinematics.

The dependence of elastic baryon form factors 
of such models on the form of kinematics has been
investigated in a previous paper \cite{bruno02}. 
With simple symmetric representations of the baryon states
a good qualitative representation was obtained with all three forms of
kinematics.
Here the investigation of Ref.~\cite{bruno02} is extended
to axial transition form factors and the pion decay widths of 
the lowest lying resonances of the nucleon and the $\Delta$.
For the nucleon, the $\Delta$ and their positive parity 
excitations, $N(1440)$ and $\Delta(1600)$, the algebraic representations 
designed in Ref.~\cite{bruno02} are employed. 
These wave functions implement the symmetries and radial shapes of
the hyperspherical 
mass operators used in~\cite{coe98}. The wave function of the lowest 
energy negative parity excitation, $N(1535)$, is constructed from the ground 
state wave function by imposing the short-range 
behavior that is implied by the centrifugal repulsion in the 
hyperspherical mass operator. Detailed quantitative description of 
the empirical features would require refinements of the baryon-states 
and/or the current matrices and very likely the consideration of 
configurations with quark-antiquark pairs.

As the pion decay of the baryons is described by the coupling
of the gradient of the pion field to the axial current of the 
baryons, the decay widths are proportional to the axial form factor.
In the case of the higher resonances, the calculated decay widths vary
significantly with the choice of kinematics. Point form kinematics
yields unrealistically small values for the pion decay widths of the 
lowest two positive parity resonances. Front and instant
form kinematics yield values, which on the average are smaller 
than half of the empirically extracted imaginary 
parts of the resonance pole positions. 

This paper is structured in the following way.
The relevant baryon kinematics and the relations of 
axial currents to pion decay widths are summarized in Section 2. Section 3 
contains the description of the quark representations of the baryon
states and a discussion of the quark kinematics used 
in the construction of axial current densities. Section 4 contains a
comparison of the axial transition form factors and pion decay widths
that are calculated with the three different forms of 
kinematics. Section 6 contains a concluding discussion.

\section{Axial Transition Form Factors and Pion Decay Widths}
\subsection{Baryon Kinematics}

The axial transition form factors are invariant matrix elements of 
the axial currents. They are functions of the invariant velocity transfer,
\beq
\eta={1\over 4}(v_f-v_a)^2\, ,
\label{eta}
\eeq
where $v_a$ and $v_f$ are the 4-velocities of the initial and final
baryons. The variable $\eta$ is related to the invariant 4-momentum
transfer as
\beq
Q^2 = (M_f v_f -M_a v_a)^2 = 4 M_f M_a \eta - (M_f - M_a)^2\, .
\label{Q2}
\eeq
For pionic decay, $N^*\to N+\pi$, the momentum transfer is $Q^2 = - m_\pi^2$ and
\beq
\eta_\pi={(M_a-M_f)^2-m_\pi^2\over 4M_fM_a}\; .
\eeq
The covariant axial current densities $A^\mu(x)$ considered here are defined by
chirally covariant Dirac spinor and Rarita-Schwinger spinor matrices,
\beq
\A_{\ohalf,\ohalf}^\mu:= \gamma^\mu \gamma_5\; , \qquad 
\A_{\ohalf,\thalf}^{\mu \nu}:=\one \,g^{\mu\nu} \, ,
\eeq
for $\ohalf,\ohalf$ and $\ohalf,\thalf$ transitions. These matrices are 
related to the spin representations of the axial current density 
operator, $A^\mu(x)$, by boost transformations and projections to definite 
intrinsic parity:
\begin{eqnarray}
&& \langle M_f,v_f, \ohalf^+ \vert A^\mu_\alpha(0) \vert\, \thalf^+,v_a,M_a\rangle
= G_A^{\Delta3/2^+}\!(Q^2)\, \bar u(v_f)\,u^\mu(v_a)\,\chi^\dagger
\chi_\alpha
\, , \label{GAdelta}\\
&&\langle M_f,v_f,\ohalf^+\vert A^\mu_a(0)\vert\,\ohalf^+,v_a,M_a\rangle
= G_A^{N1/2^+}\!(Q^2)\, \bar u(v_f)\,\, \gamma^\mu\gamma_5 u(v_a)\,
\chi^\dagger\tau_\alpha\chi\, ,\label{GAroper} \\
&&\langle M_f,v_f, \ohalf^+\vert A^\mu_a(0)\vert\, \ohalf^-,v_a,M_a\rangle
=G_A^{N1/2^-}\!(Q^2)\,  \bar u(v_f)\, \gamma^\mu u(v_a)\,
\chi^\dagger
\tau_\alpha\chi\, \label{GAneg}. 
\label{GA}
\end{eqnarray}
Here $\chi$ and $\chi_a$ denote isospin amplitudes for isospin $1/2$ and 
$3/2$, and $G_A(Q^2)$ is an invariant form factor. Explicit expressions for 
the spinor amplitudes $u(v)$ and $u^\mu(v)$ are listed in Appendix~\ref{app:spin}.

With canonical spin the spin quantization axis is in 
the direction of the velocities. The choice of frame is 
a matter of convenience. For instance \cite{bruno02}:
\beq
v_f=\left\{\sqrt{1+\eta},0,0,\sqrt{\eta}\right\},\quad
v_a=\left\{\sqrt{1+\eta},0,0,-\sqrt{\eta}\right\}\, .
\label{pointvel}
\eeq 
With this choice the canonical-spin matrices are for $(\ohalf,\ohalf)$ transitions
\beqa
\bra{\ohalf^+}A(0)\ket{\ohalf^+}&:= &G_A(Q^2)
\left\{0, \sqrt{1+\eta}\,\sigma_\perp,  \sigma_z\right\}\nonumber \\
\bra{\ohalf^+}A(0)\ket{\ohalf^-}&:=& G_A(Q^2)
\left\{1,\imath\sqrt{\eta} (\hat z\times \vec \sigma),0\right\}\; ,
\label{AX1}
\eeqa
and for $(\ohalf^+,\thalf^+)$: 
\beqa
&&\bra{\ohalf^+,\sigma'}\sqrt{\ohalf}[A_x(0)
\mp\imath A_y(0)]\ket{\thalf^+,\sigma}:=-\sqrt{\tturd}
(\ohalf, 1, \sigma',\pm1|\thalf \sigma) G_A(Q^2)\, , \nonumber \\
&&\bra{\ohalf^+,\sigma'}A_z(0)\ket{\thalf^+,\sigma}:=-\sqrt{\tturd}
(\ohalf, 1, \sigma',0|\thalf \sigma) G_A(Q^2)\; ,\qquad  A^0(0)=0\, .
\label{AX2}
\eeqa

Under the longitudinal Lorentz transformations,
\beqa
&&v_f=\{\sqrt{1+\eta} \cosh\theta+\sqrt{\eta}\sinh\theta,0,0,
\sqrt{\eta}\cosh\theta+\sqrt{1+\eta}\sinh\theta\}\, ,\nonumber \\
&&v_a=\{\sqrt{1+\eta}\cosh\theta-\sqrt{\eta}\sinh\theta,0,0,
-\sqrt{\eta}\cosh\theta+\sqrt{1+\eta}\sinh\theta\}\, .
\label{vavfi}
\eeqa
the longitudinal components of the current matrices are covariant and 
transverse components are invariant. The Wigner rotations of the 
spins are the identity .

For the null-plane spin representation, with the null vector
$n:=\{-1,0,0,1\}$, and
\beq
v_f:=\left\{\sqrt{1+\eta},\sqrt{\eta},0,0\right\}\;, \qquad 
v_a=\left\{\sqrt{1+\eta},-\sqrt{\eta},0,0\right\}\, ,
\eeq
the transverse components of the momentum transfer: 
\beq
Q=(M_f-M_a){v_f+v_a\over 2}+{M_f+M_a\over 2}(v_f-v_a)
\eeq
are
\beq
Q_\perp= \sqrt{\eta} \{ M_a + M_f , 0\}\;, 
\eeq
and
\beq
Q^2={4M_fM_a\over (M_f+M_a)^2}\, Q_\perp^2 -(M_f-M_a)^2\, .
\eeq
This is a generalization of the standard $n\cdot Q=0$ convention
to inelastic transitions where the momentum transfer may be time-like.

The null-plane spin matrices are then related to the form factors by
\beq
\bra{\ohalf^+} n\cdot A(0)\ket{\ohalf^+}= \sqrt{1+\eta}\,G_A(Q^2) \sigma_z \; ,\qquad
\bra{\ohalf^+}n\cdot A(0)\ket{\ohalf^-}=\sqrt{1+\eta}\, G_A(Q^2) \; ,
\eeq
and
\beq
\bra{\ohalf^+,\ohalf}n\cdot A(0)\ket{\thalf^+,\ohalf}={2\over 3}
\left({M_f+M_a\over 2\sqrt{M_aM_f}}\right)
\sqrt{1+\eta}\, G_A(Q^2)\; .
\eeq

\subsection{ The pion decay widths}

The pion coupling to the baryons is taken to have the 
chiral pseudovector form:
\beq
\L(x) = - {1\over 2 f_\pi} \partial_\mu  \pi_\alpha(x) \cdot 
A_\alpha^\mu(x)\, ,
\label{picoupling}
\eeq
where $f_\pi$ is the pion decay constant ($f_\pi$ =93 MeV) and 
$A_\mu$ is the axial current of the hadron. 

The perturbation $\M'$ of the mass operator is given by $H'$ in the ``rest frame''
of the excited baryon, $\vec p\,^*=0$. The energies of the initial baryon, 
the nucleon and the pion are then
\beq
 E^*=M^*\; , \qquad E= \sqrt{M^2+\vec q^2} \;, \qquad 
\omega:=\sqrt{m_\pi^2+\vec q^2}\, .
\eeq
This implies that the perturbation $\M'$ of the mass operator $\M$ is represented by
\beq
\bra{ \vec q, \sigma_N, \tau_N,\tau_\pi}\M'\ket{ \P,I,j,\sigma, \tau}:=
{1\over \sqrt{(2\pi)^3}}\sqrt{{M\over 2 \omega E}}
{\imath \over\sqrt{ 2f_\pi}}
\bra{\vec q, \sigma_N,\tau_N}q_\mu A^\mu_{\tau_\pi}(0)\ket{0,\P,I,j,\sigma, \tau}\, ,
\eeq
where $\P$, $I$ and $j$ are the parity, isospin and spin of the excited 
baryon, and $q:=\{\omega,-\vec q\}$.

As a consequence the decay width, $\Gamma$: 
\beqa
\Gamma &:=&  {1\over 2I +1}\sum_\tau {1\over 2j+1} \sum_\sigma 
\sum_{\sigma_N ,\tau_N} \sum_{\tau_\pi}\cr\cr
&\times&{1\over 2\pi}\int d^3 q 
\delta\left(\sqrt{M^2+q^2}+\sqrt{m_\pi^2+q^2}-M^*\right)
|\bra{ \vec q, \sigma_N, \tau_N,\tau_\pi}\M'\ket{ \P,I,j,\sigma, \tau}|^2\, ,
\eeqa
is proportional to $(G_A(-m_\pi^2)/f_\pi)^2$.
For the resonances considered here explicit expressions are~\cite{Cheng}:
\beqa
&&\Gamma_{\Delta(3/2 ^+)}= {1\over 48\pi} \; \bigg|{G_A^{\Delta3/2 ^+}
 \!\!(-m_\pi^2) \over  f_\pi} \bigg|^2 \;
{\sqrt{q_\pi^ 2 + M^2}+M\over  M^*}
q_\pi^3\, , \label{wdelta} \\
&&\Gamma_{N(1/2 ^+)} = {3\over 16\pi} 
\; \bigg| { G_A^{N 1/2 ^+}\!\!(-m_\pi^2) \over
  f_\pi} \bigg|^2 \;
{\sqrt{q_\pi^ 2 + M^2}-M\over M^*}
(M^*+M)^2 \;q_\pi\, , \label{wroper}\\
&&\Gamma_{N(1/2 ^-)} = {3\over 16\pi}
\bigg|{G_A^{N 1/2 ^-}\!\!(-m_\pi^2) \over  f_\pi}
 \bigg|^2 \;
{\sqrt{q_\pi^ 2 + M^2}+M\over  M^*}
(M^*-M)^2 \; q_\pi\, ,
\label{w1535}
\eeqa
where 
\beq
q_\pi = { \sqrt{ [M^{*2}-(M-m_\pi)^2] [M^{*2}-(M+m_\pi)^2]}\over 2 M^*} \; . 
\label{qpi}
\eeq

\section{Representations of Baryon States and Quark kinematics}

\subsection{Representations of Baryon States }

The baryon states formed of 3 confined quarks with total four-momentum 
$P=M v$, may be represented by functions of the form
\beq
\Psi_{M,j,v_a,\sigma}(\vec v; \vec k_1,\vec k_2, \vec k_3;
\sigma_1,\sigma_2,\sigma_3)=
\phi_\sigma(\vec k_1,\vec k_2,\vec k_3;\sigma_1,\sigma_2,\sigma_3)
\delta^{(3)}(\vec v-\vec v_a)\; .
\eeq
Here $\vec k_i$ and $\sigma_i$ are constituent momenta and spin variables.
The flavor and color variables are suppressed. The norm of the wave function 
$\phi_\sigma$ is specified by
\begin{equation}
\|\phi_\sigma\|^2=\sum_{\sigma_1,\sigma_2,\sigma_3}
\int d^3 k_1\int d^3 k_2\int d^3 k_3 \delta\left(\vec k_1+\vec k_2+\vec k_3\right)
 \delta_{\sigma_1+\sigma_2+\sigma_3,\sigma}\,
\left\vert  \phi_\sigma(\vec k_1,\vec k_3,\vec k_3;
\sigma_1,\sigma_2,\sigma_3)\right\vert^2\; ,
\label{norm}
\end{equation}
which implies that
\beq
(\Psi_{M,v_f,\sigma'},\Psi_{M,v_a,\sigma})=
\delta^{(3)}(\vec v_f-\vec v_a)
\delta_{\sigma',\sigma}.
\eeq
Under Poincar\'e transformations the velocity $v$ transforms 
as a four-vector, while the total spin operator $\vec j$ undergoes 
Wigner rotations, ${\cal R}_W(\Lambda,v):=B^{-1}(\Lambda v)\Lambda B(v)$,
\beq
U^\dagger(\Lambda,d)\, v\, U(\Lambda,d)= \Lambda v \; ,\qquad v^2=-1\;,\qquad
U^\dagger(\Lambda,d)\, \vec j\, U(\Lambda,d)= 
{\cal R}_W(\Lambda,v) \vec j\; .
\eeq
Here the boost $B(v)$ is the rotationless Lorentz transformation,
which satisfies the defining relation $B(v)\{1,0,0,0\}=v$.

The constituent momenta and spins undergo the same Wigner rotations:
\beq
U^\dagger(\Lambda)\, \vec k_i\, U(\Lambda)={\cal R}_W(\Lambda,v)\, 
\vec k_i\; ,\qquad
U^\dagger(\Lambda)\, \vec j_i\, U(\Lambda)={\cal R}_W(\Lambda,v)\, 
\vec j_i\; .
\label{LL}
\eeq
The Poincar\'e covariance of the bound-state wave function 
$\phi_\sigma$, is realized by its covariance under rotations, 
invariance under translations and its independence of the 
velocity $v$. 

The constraint $\sum_i\vec k_i =0$ is conveniently implemented 
by the definition of Jacobi momenta,
\beq
\vec{\kappa}:=\sqrt{2\over 3}\left(\vec{k}_1 - 
{\vec{k}_2+\vec{k}_3\over2}\right)\, , \qquad
\vec{q}:=\sqrt{1\over2}(\vec{k}_2-\vec{k}_3) \, ,
\eeq
which implies
\beq
\vec k_1^2+\vec k_2^2+\vec k_3^2= \vec \kappa^2+\vec q^{\,2}\;.
\eeq

The wave functions representing the positive parity states considered
are products of the permutation symmetric spin-isospin amplitudes 
$\chi_{j,\sigma,\tau}$, and functions 
of $\sP:= \sqrt{2(\vec \kappa^2+\vec q^2)}$, \cite{bruno02}
\beq
\phi_{n,j,\sigma,\tau}(\vec \kappa,\vec q)= \chi_{j,\sigma,\tau}\; \varphi_n(\sP)\; ,
\eeq
with
\beq
\varphi_0(\sP)= {\cal N} \left( 1+ {\sP^2\over 4 b^2}\right)^{-a}\; ,
\label{ground}
\eeq
and
\beq
\varphi_1(\sP) = \varphi_0(\sP) \left(A(a)  
+B(a)\left[  - {12 a\over  (1+{\sP^2\over 4 b^2})}
+ {a(a+1) \over b^2}   {\sP^2 \over  (1+{\sP^2\over 4 b^2})^2}  \right]\right)\;.
\label{roperexpl}
\eeq
The factor ${\cal N}$ is a normalization constant. The exponent 
$a$ and scale $b$ are parameters. The constants $A(a)$ and $B(a)$ are determined, 
by the orthonormality condition 
\beq
  \int d^3{\kappa} \; d^3{q} \; \varphi_i(\sP)
\varphi_j(\sP) = \delta_{ij}. 
\label{norm1}
\eeq

\begin{table}[b]
\caption{Values of the parameters of the ground 
state wave function used for the three different 
forms of kinematics. The corresponding matter radii
$r_0$ are listed in the last column.\label{parameters}}
\begin{ruledtabular}
\begin{tabular}{lcccc}
             &  $m_q$ (MeV)   &  $b$ (MeV)  &  $a$ &$r_0$ (fm) \\  
\hline
instant form &  140           &  600        &   6  & 0.63  \\
point form   &  350           &  640        &  9/4 & 0.19 \\
front form   &  250           &  500        &   4  & 0.55 \\ 
\end{tabular}
\end{ruledtabular}
\end{table}

The representation of the $1/2^-$ resonance $N(1535)$ involves 
the permutation symmetric bilinear combination of 
the mixed-symmetry orbital functions of
$\{\vec \kappa\, ,\, \vec q\}$ and the mixed-symmetry spin-isospin amplitudes 
$\{\chi_{\ohalf+.\sigma,\tau}, \chi_{\ohalf -,\sigma,\tau}\}$:
\beq
\phi_{\ohalf^-,\sigma,\tau}(\vec \kappa,\vec q) = 
{1\over 2}\sum_{ms}(1,\ohalf,m,s|\ohalf, s_3)
 \bigg\{ \kappa \; Y_{1m}\;(\hat \kappa)\chi_{\ohalf+,\sigma,\tau} 
 + q\; Y_{1m}(\hat q)\; \chi_{\ohalf-,\sigma,\tau}\bigg\}
\varphi_2(\sP)\, .
\label{1/2minus}
\eeq
The parameters determined in Ref.~\cite{bruno02} provide a qualitative
 fit to the electromagnetic form factors of the nucleon 
with each form of kinematics. They are listed in 
Table~\ref{parameters}.
For the radial wave function $\varphi_2(\sP)$ 
the following form is employed:
\beq
\varphi_2 (\sP) = {C\varphi_0(\sP)\over\left(1+\sP^2/ 4 b^2\right)}\, ,
\label{pstate}
\eeq
where $C$ is a normalization constant. The asymptotic behavior for large $\sP$
implements the short-range behavior of the Fourier transform, $\varphi(\sR)$, 
implied by the centrifugal repulsion~\cite{coe98}.

\subsection{Quark Kinematics}

The kinematic quark-current matrices are functions of quark momenta that are related
by kinematic boosts to the constituent momenta $\vec k_i$.
With point form kinematics all Lorentz transformations
are kinematic, and the velocities are kinematic variables. The 
canonical boost $B_c(v)$ relates the quark momentum $p_i$ to the 
constituent momentum $\vec k_i$,
\beq
p_i:= B(v)\left\{\sqrt{ m_q^2+\vec k_i^2},\vec k_i\right\}\, .
\eeq
 
Kinematic null-plane quark momenta $p_i^+$ and $p_{i\perp}$ are related to
the momentum fractions $\xi_i$ and constituent momenta $k_{i\perp}$
by 
\beq
p_i^+= \xi_i P^+ \qquad \mbox{and}\qquad p_{i\perp}= k_{i\perp}+\xi_i P_\perp ,
\eeq
and the null-plane spins are related to Dirac spinors the spinor representations
of null-plane boosts~\cite{Chung,coe92}.

With instant form kinematics the boost parameters are the 
initial and final momenta, $\vec P_a$ and $\vec P_f$. These are 
related kinematically only if $\vec P_f=-\vec P_a=\vec Q /2$.
This requirement is satisfied by the 
Lorentz transformation~(\ref{vavfi}) with~\cite{bruno02}:
\beq
\theta := \tanh^{-1} \bigg[ {M_a-M_f\over M_f+M_a}\sqrt{{\eta\over 1+\eta}}
 \bigg ]\, .
\eeq
The
relation between the three momentum transfer $\vec Q$ and the
four-momentum transfer is then
\beq
\vec Q^2 = Q^2-{[(P_f+P_a)\cdot Q]^2\over (P_f+P_a)^2}
=Q^2+{(M_f^2-M_a^2)^2\over Q^2+2(M_f^2+M_a^2)}\,.
\label{Qs}
\eeq 
The kinematic quark momenta $p_i$ and velocities, $v_i:= p_i/m_q$,
 are related to the constituent momenta $\vec k_i$
by canonical boosts parameterized by $\vec P$,
\beq
p_i:= B_c(\vec P/M_0)\left\{\sqrt{m_q^2+\vec k_i^2},\vec k_i\right\}\;
, \qquad 
M_0:= \sum_i \sqrt{m_q^2+\vec k_i^2}\; .
\eeq

Axial current operators are generated by the dynamics from 
single-quark current matrices
$
\bar u_c(p_i')\gamma_\perp\gamma_5 u_c(p_i)
$
with point and instant kinematics, and
$
\bar u_f(p_i')\gamma^+\gamma_5u_f(p_i)
$
with null-plane kinematics.

\section{Axial Baryon Form Factors and Pion Decay Widths }

\subsection{The axial form factor of the nucleon}

\begin{figure}[tb]
\vspace{25pt}
\begin{center}
\mbox{\epsfig{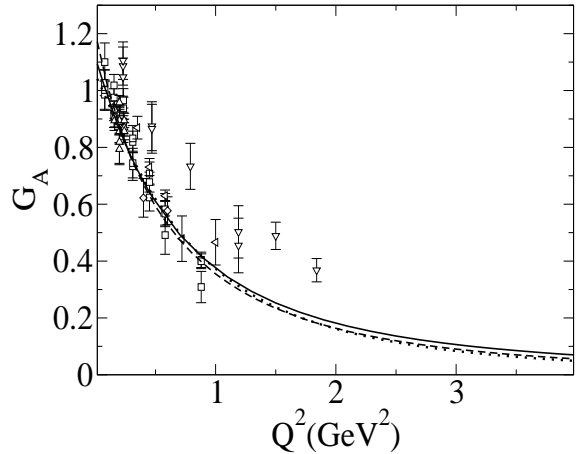}}
\end{center}
\caption{Axial form factor of the nucleon. Solid, dotted and dashed lines 
correspond to the instant, point and front forms respectively
(from \protect\cite{bruno02}). 
\label{GANUC}}
\end{figure}

The axial form factors as obtained in the three forms of 
kinematics and with the wave functions defined above have 
been given in Ref.~\cite{bruno02}. These results are compared in Fig.~\ref{GANUC} to
the corresponding empirical values given in Ref.~\cite{ga}.
All the calculated form factors agree, within the uncertainty 
limits, with the extant data, with little dependence on the 
choice of kinematics, due to the fact that different spatial wave 
functions are employed with the different kinematic currents. The corresponding 
values for the axial vector coupling constant of the nucleon 
$G_A(0)$ are 1.1 with instant and point form kinematics, and
1.2 with front form kinematics. These values are thus smaller by
$5-15\%$ than the empirical value 1.267. In comparison the static
quark model value 5/3 exceeds the empirical value. 

\subsection{The $\Delta(1232)\rightarrow N$ axial transition form factor}

\begin{figure}[bt]
\vspace{25pt}
\begin{center}
\mbox{\epsfig{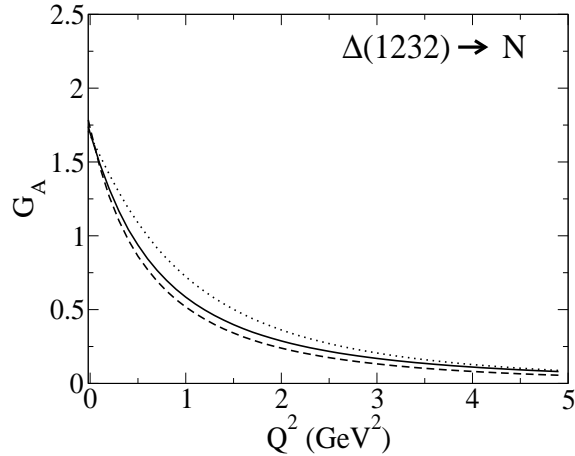}}
\end{center}
\caption{Axial form factor for the $N\to \Delta(1232)$
transition. Solid, dotted and dashed lines correspond to the 
instant, point and front forms respectively.\label{GA1232}}
\end{figure}

The calculated values for the $\Delta(1232)\rightarrow N$ 
transition form factor are shown in Fig.~\ref{GA1232}. The 
only empirical information on this form factor is that 
obtained by neutrino-induced reactions on the deuteron in 
Ref.~\cite{Kitagaki}. The value for $G_A^\Delta(0)$ extracted
from this experiment is $G_A^\Delta(0)/\sqrt{2} = 2.4 \pm 0.25$ 
\cite{Hemmert}. 
The present values for $G_A^\Delta(-m_\pi^2)$ that are obtained in 
instant, point and front form kinematics are $G_A^\Delta(-m_\pi^2)= 1.74$, 
$G_A^\Delta(-m_\pi^2)= 1.70$ and $G_A^\Delta(-m_\pi^2)= 1.79$ thus clearly 
too small and also smaller than the semirelativistic
quark model value 2.12 obtained in Ref.~\cite{Hemmert}. 

For comparison the static quark model value for $G_A^\Delta(-m_\pi^2)$ 
is $2\sqrt{2}$. It is worth noting than the experimental value for 
$G_A^\Delta(-m_\pi^2)$ leads to a value for the pion decay 
constant which is too small by 30 \%.

The results obtained are consistent with the assumption of quark-antiquark 
components in addition to the three-quark structure of the $\Delta(1232)$, 
which is supported by the observation that the ratios
of the experimental pionic decay widths of $\Delta\to N \pi$, 
$\Sigma^* \to  \Sigma\pi$, and $\Xi^*\to \Xi \pi$ do not follow 
the simple 3-quark expectation $9:4:1$ but are $12:4:1$, 
$\Delta(1232)$ ~\cite{riska03}.

In instant and front form $G_A^\Delta$ has very similar
dependence on momentum transfer as the calculated corresponding 
elastic axial form factor of the nucleon. In contrast the 
calculated $G_A^\Delta$ in point form falls at a considerably
slower rate than the corresponding elastic axial form factor of 
the nucleon.

\subsection{The $N(1440)\rightarrow N$ axial transition form factor}

\begin{figure}[bt]
\vspace{25pt}
\begin{center}
\mbox{\epsfig{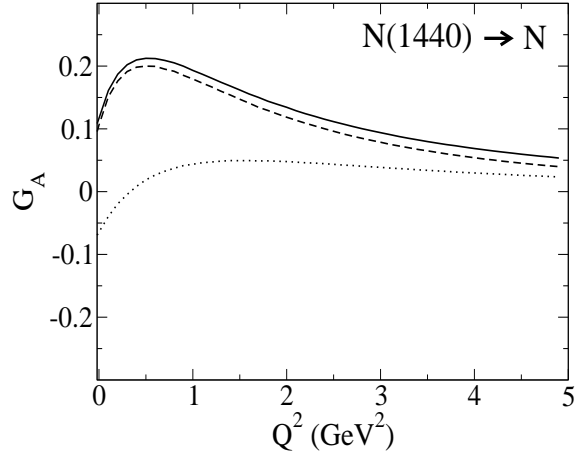}}
\end{center}
\caption{Axial form factor of the $N(1440)\to N$ transition. 
Solid, dotted and dashed lines 
correspond to the instant, point and front forms respectively. \label{GA1440}}
\end{figure}

The calculated $N(1440)\rightarrow N$ axial transition
form factors are shown in Fig.~\ref{GA1440}. 
The form factors are qualitatively similar in the three
forms, in all cases they change sign near $Q^2=0$. In instant and 
front form kinematics the sign change takes place in the timelike 
region ($Q^2<0$). The calculated values at the pion 
point are in all forms of kinematics smaller by a 
factor $\sim$ 4 than the corresponding values 
extracted from experiment that are shown in Table~\ref{pionp}. 

\subsection{The $N(1535)\rightarrow N$ axial transition form factor}

\begin{figure}[t]
\vspace{25pt}
\begin{center}
\mbox{\epsfig{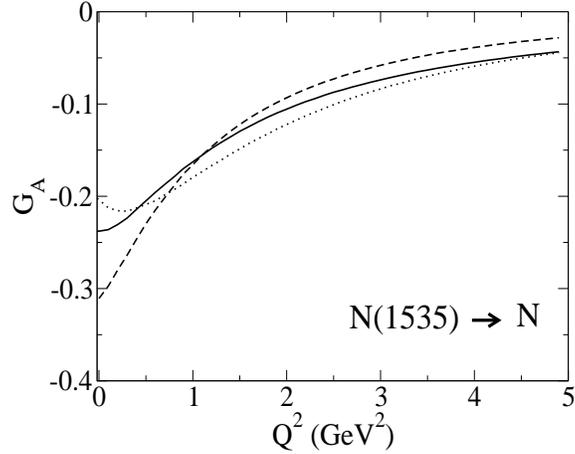}}
\end{center}
\caption{Axial form factor for $N\to N(1535)$ transition. 
Solid, dotted and dashed lines correspond to the instant, 
point and front forms respectively.\label{GA1535}}
\end{figure}

In Fig.~\ref{GA1535} the $N(1535)\rightarrow N$ 
axial transition form factor is shown. The detailed evaluation is 
given in Appendix~\ref{app1535}. The $Q^2$ behavior of the 
form factor with the three forms is similar above $Q^2>2$ GeV$^2$, however
the value at the pion point, $G_A^{N(1535)}(-m_\pi^2)$, differs 
between the forms. All forms of kinematics give values that are close to the
range of the values extracted from experiment (Table~\ref{pionp}).

\subsection{The $\Delta(1600)\rightarrow N$ axial transition form factor}

\begin{figure}[t]
\vspace{25pt}
\begin{center}
\mbox{\epsfig{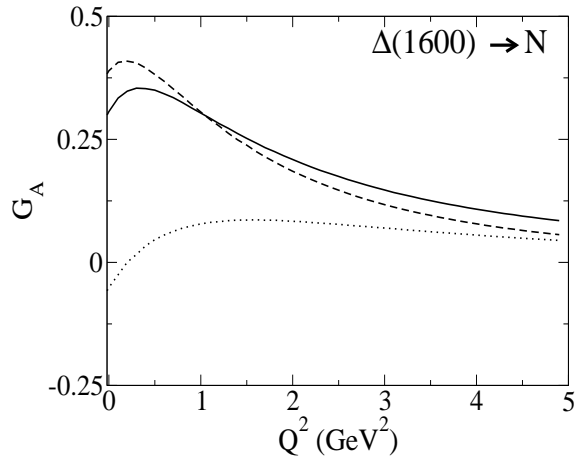}}
\end{center}
\caption{Axial form factor for the nucleon to Delta(1600) transition. 
Solid, dotted and dashed lines 
correspond to the instant, point and front forms respectively. \label{GA1600}}
\end{figure}

The axial form factors calculated with different kinematic currents 
are shown in Fig.~\ref{GA1600}. The differences are similar to those 
found for the $N(1440)\to N$ transition although the difference between 
instant and point form are more pronounced. At the pion point point form 
kinematics yields an unrealistically small value, (Table~\ref{pionp}).

\subsection{ Pion Decay Widths}

The value of the axial form factor at the pion point, 
$G_A(Q^2=-m_\pi^2)$, is related through expressions 
(\ref{wdelta}), (\ref{wroper}) and (\ref{w1535}), to 
the pion decay width of the corresponding resonances.

The calculated values for the pion decay widths are
listed in Table~\ref{decayw} along with the empirical
ranges for those values. The results in this table
reveal that none of the form of kinematics 
is able to get values commensurable with the empirical ones.
Instant and front form provide a comparable picture 
of the decay widths of these baryon resonances, while 
point form fails more abruptly than the others, in particular
in the case of positive parity resonances.
This failure of the point form of kinematics to provide a
plausible picture of the pion decay widths has 
already been reported in Ref.~\cite{melde}. 

\begin{table}[hb]
\caption{Axial form factors at the pion point, $|G_A(-m_\pi^2)|$.
The PHEN column contains the corresponding axial coupling constants 
that have been obtained phenomenologically from 
the imaginary parts of the corresponding resonance pole
positions \cite{PDG} using Eqs.~(\ref{wdelta}) and~(\ref{w1535}).}
\label{pionp}
\begin{ruledtabular}
\begin{tabular}{lcccc}
              &  instant       &  point      &  front       & PHEN  \\
\hline
$\Delta(1232)$&    1.74        &  1.70       &  1.79        & 2.67 \\
$N(1440)$     &    0.11        &  0.07       &  0.10   & [0.38$-$0.41]\\
$N(1535)$     &    0.24        &  0.20       &  0.31   & [0.20$-$0.25]\\
$\Delta(1600)$ &   0.30        &  0.06       &  0.38   & [0.48$-$0.75]\\
\end{tabular}
\end{ruledtabular}
\end{table}

\begin{table}[hb]
\caption{Pion decay widths in MeV that correspond to 2 times 
the imaginary pole positions in~\cite{PDG}.
\label{decayw}}
\begin{ruledtabular}
\begin{tabular}{lccccc}
                    &  instant       &  point     &  front     & PHEN \protect
\cite{PDG} \\   
\hline
$\Delta(1232)$      &  42            &  41        &  45    & 100  \\
$N(1440)$           &  10            &  4         &   8    & [126$-$147]  \\
$N(1535)$           &  84            &  60        &  147    & [59$-$93]   \\ 
$\Delta(1600)$      &  12            &  0         &  19    & [30$-$75]  \\
\end{tabular}
\end{ruledtabular}
\end{table}

\section{Discussion}

The results for the calculated axial transition form factors
of the baryon resonances with the present simple representations 
show dependence on the kinematics used to generate the 
current operators. The spatial wave function was in each case 
parameterized so as to describe both the electromagnetic as well 
as the axial form factors of the nucleon (cf.~Fig.~\ref{GANUC}). 
The results also exhibit a strong dependence on the representation of the
excited states.

The calculated pion decay widths are in most cases considerably
smaller than the empirically extracted values. This suggests that 
a description of the empirical values would call for sizable
quark-antiquark components in the wave functions. Point form 
kinematics leads to unrealistically small values for the pion decay 
widths of the radial excitations of the nucleon and the $\Delta(1232)$ 
resonance. From the point of view of quark model phenomenology
for the baryons the present results indicate a slight preference for 
front form kinematics over instant form kinematics. 

\begin{acknowledgments}
B. J.-D. thanks the European Euridice network for 
support (HPRN-CT-2002-00311). Research supported in part by 
the Academy of Finland through grant 54038 and by the U.S. Department 
of Energy, Nuclear Physics Division, contract W-31-109-ENG-38.

\end{acknowledgments}

\appendix
\section {Boosts relating spin and spinor representations}
\label{app:spin}

Spin and spinor representations are related by appropriate representations
of the Lorentz boosts $B(v)$.
For spin $\ohalf$ the spinor representation of canonical boosts $B_c(v)$
and null-plane boosts $B_f(v)$ are
\beqa
S[B_c(v)]&:=& D^{\ohalf,0}[B_c(v)]\oplus D^{0,\ohalf}[B_c(v)]=
{\vec \alpha\cdot \vec v + 1 +v^0\over\sqrt{2(1+v^0)}}\\
S[B_f(v)]&:=& D^{\ohalf,0}[B_f(v)]\oplus D^{0,\ohalf}[B_f(v)]=
{\alpha_\perp\cdot  v_\perp + v^+\over\sqrt{v^+}}{1+\alpha_z\over 2}
+{1\over \sqrt{v^+}}{1-\alpha_z\over 2} \, .
\eeqa
The spinor amplitudes $u_c(v)$ and $u_f(v)$ are defined by the boosts applied to the
projections $(1+\beta)/2$ onto positive intrinsic parity,
\beq
u_c(v):=S[B_c(v)]{1+\beta\over 2}\; , \qquad u_f(v):= S[B_f(v)]{1+\beta\over 2}\; .
\eeq

For spin $\one$ the appropriate boost representations are $D{\ohalf,\ohalf}[B(v)]$
and the projections are the projections onto 4-vectors with vanishing time components.
It follows that the spin-$\one$ spinor amplitudes are 4-vectors orthogonal to 
the velocity. With canonical boosts the result is
\beqa
\epsilon_\lambda(p) &:=&  e_\lambda + {e_\lambda\cdot p \over m+\omega}
 \left[ {p\over m}+n_c\right] \, ,  
\qquad \lambda=0\pm 1 
\label{poli}
\eeqa
with $n_c:=\{1,0,0,0\}$ and
\beq
e_\lambda = - \lambda { \hat{e}_x +\lambda i \hat{e}_y \over \sqrt{2}}
 \qquad \lambda=\pm1 ;  \qquad e_0=\hat{e}_z \, .
\eeq
With null-plane boosts the amplitudes are
\beqa
 \epsilon_\lambda(p) &:=&  e_\lambda - {e_\lambda\cdot p \over n\cdot p} n 
 \, ,  \qquad \lambda=0\pm 1  \nonumber \\
 \epsilon_0(p) &=& {m\over n\cdot p} \left[ n -{(n\cdot p) p\over p^2}
 \right] \, ,
\label{polf}
\eeqa
where $n:=\{-1,0,0,1\}$.
The Rarita-Schwinger vector-spinor amplitudes of spin $\thalf$ baryons
in Eq.~(\ref{GAdelta}), are
\beq
u^\mu_\sigma(v) = \sum_{\lambda,\sigma'}
\epsilon_\lambda^\mu(v) u_\sigma(v)(1,\ohalf,\lambda,\sigma'\vert \thalf,\sigma) \, . 
\eeq

\section{$N(1535)\to N$ axial transition form factor}
\label{app1535}

When evaluating the transition form factor, $N(1535)\to N$, 
regardless of the form of kinematics, the following matrix 
elements need to be evaluated:
\beqa
\bra{p;\ohalf}  \,{\cal X\,F\,S }\, \ket{N(1535);\ohalf}&=&
{1\over \sqrt{2}}\sum_{ms} \clg{1}{\ohalf}{m}{s}{\ohalf}{\ohalf}
\;\bra{p,\ohalf}\, {\cal X\,F\,S} \,   \ket{N_{(1535)}^{ms}}\,,
\eeqa
where ${\cal X\,F\,S }$ is, in general, an operator that acts 
on spatial, flavor and spin degrees of freedom. The operator 
entering the actual evaluation is a Wigner or Melosh 
rotated axial current operator. Its flavor part 
is simply: ${\cal F}=\tau_z$.
The flavor matrix elements can be evaluated leaving the spin and 
spatial matrix elements, which do depend on the form of kinematics. 
We can write the wave functions explicitly in terms of 
their spin, flavor (some flavor indexes have been 
omitted for clarity) and spatial components: 
\beqa
\bra{p,\ohalf}\, {\cal X\,F\,S} \,  \ket{N_{(1535)}^{ms}} &=&
[3]_x \, [3]_{FS,S}^{1/2}
\,{\cal  X\,F\,S} \, 
{1\over \sqrt{2}} \left( [21]_{x,S}^m [21]_{FS,S}^s 
                       + [21]_{x,A}^m [21]_{FS,A}^s \right) \\
&=&
 \left([3]_x \,  {\cal X}\, [21]_{x,S}^m\; [3]_{FS,S}^{1/2} \,{\cal  F\,S} \,
 [21]_{FS,S}^s 
 +[3]_x \,  {\cal X}\, [21]_{x,A}^m \; [3]_{FS,S}^{1/2} 
  \,{\cal  F\,S} \, [21]_{FS,A}^s \right)\nonumber 
\label{appBeq2}
\eeqa
where the spin-flavor symmetric and mixed-symmetric wave functions are,
\beqa
\chi_{\sigma,\tau} \equiv \left[3\right]_{FS,S}^{\sigma,\tau} &=& {1\over 
\sqrt{2}} \left( [21]_{F,S}^\tau [21]_{S,S}^\sigma + 
[21]_{F,A}^\tau [21]_{S,A}^\sigma \right)\, , \nonumber \\
\chi_{+,\sigma,\tau} \equiv \left[21\right]_{FS,S}^{\sigma,\tau} &=& {1\over \sqrt{2}} 
\left( [21]_{F,S}^\tau [21]_{S,S}^\sigma - [21]_{F,A}^\tau [21]_{S,A}^\sigma \right)\, , \nonumber \\
\chi_{-,\sigma,\tau} \equiv \left[21\right]_{FS,A}^{\sigma,\tau} &=& {1\over \sqrt{2}} 
\left( [21]_{F,S}^\tau [21]_{S,A}^\sigma + [21]_{F,A}^\tau [21]_{S,S}^\sigma \right) \, .
\label{fswf}
\eeqa
with the usual,
\beqa
[3]_{S(T),S}^{l_z}&=& \sum_{abcm} 
(\ohalf,\ohalf,a,b|1,m) 
(\ohalf,1,c,m|\thalf,l_z)
\;\ket{\ohalf, a}\ket{\ohalf, b} \ket{\ohalf, c}\, , \nonumber \\
 \left[21\right]_{S(T),S}^{l_z} &=& \sum_{abcm}
 (\ohalf,\ohalf,a,b|1,m) 
(\ohalf,1,c,m|\ohalf,l_z)
\;\ket{\ohalf, a}\ket{\ohalf, b} \ket{\ohalf, c}\, , \nonumber \\
 \left[21\right]_{S(T),A}^{l_z} &=& \sum_{ab}
 (\ohalf,\ohalf,a,b|0,0) 
\;\ket{\ohalf, a}\ket{\ohalf, b}\ket{\ohalf, l_z} \,.
\eeqa

The spin-flavor matrix elements can be split into flavor and spin parts. 
The only nonzero flavor matrix elements are: 
$[21]_{F,S} \,\tau_z\, [21]_{F,S}=-1/3$ and $[21]_{F,A} \,\tau_z\, [21]_{F,A}=1$.
Then the mixed-symmetric to symmetric part of Eq.~(\ref{appBeq2}) reads:
\beqa
 [3]_{FS,S}^{1/2} \,{\cal F\,S}\,  [21]_{FS,S}^s &=& 
 {1\over \sqrt{2}} \left( [21]_{F,S} [21]_{S,S}^{1/2} + [21]_{F,A} [21]_{S,A}^{1/2}
 \right) 
\,{\cal F\,S}\,
 {1\over \sqrt{2}} \left( [21]_{F,S} [21]_{S,S}^s 
- [21]_{F,A} [21]_{S,A}^s \right) \nonumber \\
&=&
 {1\over 2} \bigg( 
 [21]_{F,S} \,{\cal F}\,  [21]_{F,S}\,  [21]_{S,S}^{1/2} \,{\cal S}\, [21]_{S,S}^s -
 [21]_{F,A}  \,{\cal F} \,  [21]_{F,A}\,  
[21]_{S,A}  \,{\cal S}\, [21]_{S,A}^s \bigg)\nonumber \\
&=&
{1\over 2} \bigg[-{1\over3} [21]_{S,S}^{1/2}\,{\cal S}\, [21]_{S,S}^s 
                 - [21]_{S,A}^{1/2}\,{\cal S}\, [21]_{S,A}^s  \bigg]
\eeqa
while the mixed-antisymmetric to symmetric has the form,
\beqa
 [3]_{FS,S}^{1/2} \,{\cal F\,S}\,  [21]_{FS,A}^s &=& 
 {1\over \sqrt{2}} \left( [21]_{F,S} [21]_{S,S}^{1/2} 
+ [21]_{F,A} [21]_{S,A}^{1/2} \right) 
\,{\cal F\,S}\,
 {1\over \sqrt{2}} \left( [21]_{F,S} [21]_{S,A}^s 
+ [21]_{F,A} [21]_{S,S}^s \right) \nonumber \\
&=&
 {1\over 2} \bigg[   [21]_{F,S} \,{\cal F}\, [21]_{F,S}  
\; [21]_{S,S}^{1/2}\,{\cal S}\, [21]_{S,A}^s 
                   + [21]_{F,A} \,{\cal F}\, [21]_{F,A}  
\; [21]_{S,A}^{1/2}\,{\cal S}\, [21]_{S,S}^s  \bigg]\nonumber \\
&=&
{1\over 2} \bigg[-{1\over3} [21]_{S,S}^{1/2}\,{\cal S}\, [21]_{S,A}^s 
                 + [21]_{S,A}^{1/2}\,{\cal S}\, [21]_{S,S}^s  \bigg] \, .
\eeqa
giving
\beqa
\bra{p,S}\, {\cal X\,F\,S} \,  \ket{ N_{(1535)}^{ms}}    &=&
{1\over 2\sqrt{2}} \bigg(
[3]_x \,  {\cal X}\, [21]_{x,S}^m\;
\bigg[-{1\over3} [21]_{S,S}^{1/2}\,{\cal S}\, [21]_{S,S}^s 
                 - [21]_{S,A}^{1/2}\,{\cal S}\, [21]_{S,A}^s  \bigg]
\nonumber \\
&&+
[3]_x \,  {\cal X}\,  [21]_{x,A}^m \;
\bigg[-{1\over3} [21]_{S,S}^{1/2}\,{\cal S}\, [21]_{S,A}^s 
                 + [21]_{S,A}^{1/2}\,{\cal S}\, [21]_{S,S}^s  \bigg]
\bigg) \, ,
\eeqa
and our initial matrix element reads: 
\beqa
\bra{p;\ohalf}  \,{\cal X\,Q\,S }\, \ket{N(1535),\ohalf}&=&
\sum_{ms} \clg{1}{\ohalf}{m}{s}{\ohalf}{\ohalf}
{1\over 2\sqrt{2}} \\
&&\bigg(
[3]_x \,  {\cal X}\, [21]_{x,S}^m\;
\bigg[-{1\over3} [21]_{S,S}^{1/2}\,{\cal S}\, [21]_{S,S}^s 
                 - [21]_{S,A}^{1/2}\,{\cal S}\, [21]_{S,A}^s  \bigg]
\nonumber \\
&&+
[3]_x \,  {\cal X}\,  [21]_{x,A}^m \;
\bigg[-{1\over3} [21]_{S,S}^{1/2}\,{\cal S}\, [21]_{S,A}^s 
                 + [21]_{S,A}^{1/2}\,{\cal S}\, [21]_{S,S}^s  \bigg]
\bigg)\,. \nonumber 
\eeqa
The remaining matrix elements involve integrals over the spatial 
wave functions,  
$[3]_x \,  {\cal X}\, [21]_{x,S}^m$ and $[3]_x \,  {\cal X}\,  [21]_{x,A}^m$, 
together with matrix elements involving the spins, 
$[21]_{S,A}  \,{\cal S}\, [21]_{S,A}^s$ and 
$[21]_{S,A}\,{\cal S}\, [21]_{S,S}^s$, which are different 
for the different forms of kinematics.

The spin matrix elements correspond to matrix elements
of the Wigner or Melosh rotated current,
\beqa
[21]_{S,A}^{1/2}  \,{\cal S}\, [21]_{S,A}^s &=&
3 \;[21]_{S,A}  \,  \bigg[D^{1/2 \dagger}_2 D^{1/2}_2 \bigg]
                  \;\bigg[D^{1/2 \dagger}_3 D^{1/2}_3 \bigg]
                  \;\bigg[D^{1/2 \dagger}_1 I_{0q} D^{1/2}_1  \bigg]
  \, [21]_{S,A}^s 
\eeqa
and 
\beqa
[21]_{S,A}^{1/2}  \,{\cal S}\, [21]_{S,S}^s &=&
3 \;[21]_{S,A}  \, \bigg[D^{1/2 \dagger}_2 D^{1/2}_2\bigg] 
                  \; \bigg[D^{1/2 \dagger}_3 D^{1/2}_3\bigg]
                  \; \bigg[D^{1/2 \dagger}_1 I_{0q} D^{1/2}_1 \bigg]
  \, [21]_{S,S}^s \, ,
\eeqa
where $D^{1/2}$ are the representations of the corresponding Wigner or Melosh 
rotations for canonical and null-plane spins respectively.

\end{document}